\documentclass[showpacs,a4paper,twocolumn,nofootinbib,superscriptaddress]{revtex4-2}

\usepackage{newpxmath,newpxtext}
\usepackage{graphicx}
\usepackage{epsfig}

\newcommand{\lin}{\\[7pt]}

\newcommand{\pder}[2]{\dfrac{\partial#1}{\partial#2}}
\newcommand{\pdder}[3]{\dfrac{\partial^2 #1}{\partial #2 \partial #3}}

\newcommand{\dder}[2]{\dfrac{\delta#1}{\delta#2}}
\newcommand{\pdot}[1]{\dot{\partial}_{#1}}

\newcommand{\Gd}{\mathcal{G}}

\newcommand{\R}{\mathcal{R}}
\newcommand{\de}{\mathrm{d}}

\begin{document}
	
\title{Schwarzschild-like solutions in Finsler-Randers gravity}
	
\author{A. Triantafyllopoulos}
	\email{alktrian@phys.uoa.gr}
	\affiliation{Section of Astrophysics, Astronomy and Mechanics, Department of Physics, National and Kapodistrian University of Athens, Panepistimiopolis 15784, Athens, Greece}

\author{S. Basilakos}
	\email{svasil@academyofathens.gr}
	\affiliation{Academy of Athens, Research Center for Astronomy and Applied Mathematics, Soranou Efessiou 4, 115 27 Athens, Greece}
	\affiliation{National Observatory of Athens, Lofos Nymfon, 11852, Athens, Greece}

\author{E. Kapsabelis}
	\email{manoliskapsabelis@yahoo.gr}
	\affiliation{Section of Astrophysics, Astronomy and Mechanics, Department of Physics, National and Kapodistrian University of Athens, Panepistimiopolis 15784, Athens, Greece}

\author{P. C. Stavrinos}
	\email{pstavrin@math.uoa.gr}
	\affiliation{Department of Mathematics, National and Kapodistrian University of Athens, Panepistimiopolis 15784, Athens, Greece}

\begin{abstract}
	In this work, we extend for the first time the spherically symmetric Schwarzschild and Schwarzschild-De Sitter solutions with a Finsler-Randers-type perturbation which is generated by a covector $A_\gamma$. This gives a locally anisotropic character to the metric and induces a deviation from the Riemannian models of gravity. A natural framework for this study is the Lorentz tangent bundle of a spacetime manifold. We apply the generalized field equations to the perturbed metric and derive the dynamics for the covector $A_\gamma$. Finally, we find the timelike, spacelike and null paths on the Schwarzschild-Randers spacetime, we solve the timelike ones numerically and we compare them with the classic geodesics of general relativity. The obtained solutions are new and they enrich the corresponding literature.
\end{abstract}

\pacs{04.50.-h, 04.50.Kd}

\maketitle

\section{Introduction}

In the context of research on generalized metric spaces, Finsler, Lagrange and Finsler-like geometries have played an important role in modified theories of gravity, general relativity and cosmology. In the last two decades, a development of these promising topics of research has extended the limits of general relativity and cosmology by including locally-anisotropic approaches. 
Over the last two decades field equations 
have been thoroughly studied
in the context of Finsler, Lagrange, generalized Finsler 
and Finsler-like geometries 
as well as in scalar-tensor theories.  
The basic feature of these theories is the presence of extra terms 
in the equations of motion due to the intrinsic geometrical spacetime 
anisotropy. In this framework the term ``spacetime anisotropy'' is related to the Lorentz violation feature of the geometry. 
This approach 
may provide the necessary platform in understanding one of the most
crucial problems in cosmology which is related with the underlying mechanism 
of cosmic acceleration and thus of dark energy.
There is a lot of work in the literature on ``spacetime anisotropic'' geometries
and below we briefly present some relevant works. 

Generalized Einstein field equations have been studied in the Finsler, Lagrange, generalized Finsler and Finsler-like spaces, for an osculating gravitational approach in which the second variable $y(x)$ is a tangent/ vector field \cite{asanov1982,asanov1983,Asa41} and in Finsler cosmology \cite{Stavrinos:2006rf,Kouretsis:2008ha,Stavrinos:2012ty,Kouretsis:2012ys,Chaubey:2018wph}. 
Different sets of generalized Einstein field equations were 
derived for the aforementioned spaces in the framework of a tangent bundle \cite{Pfeifer:2015tua,Hohmann:2018rpp,Hohmann:2019sni,Hohmann:2020mgs,Vacaru:2009ye,Stavrinos:2014apa,Vacaru:2005ht,Triantafyllopoulos:2018bli,Triantafyllopoulos:2020ogl} and for the momentum space on the cotangent bundle \cite{Relancio:2020rys,Relancio:2020zok,Perelman2020,CastroPerelman:2018hde,Miron:1994nvt,Miron2010}. 
Additionally, Lorentz invariance violation in Finsler / Finsler-like spacetime and in Finsler cosmology in very special relativity has also been studied 
in a large series of papers \cite{Gibbons:2007iu,Elbistan:2020mca,Kostelecky:2011qz,Edwards:2018lsn,Kouretsis:2008ha}. 
Investigations on generalized scalar-tensor theories with 
Finsler-like structure modeled on a vector bundle with two-internal 
fibers have also been done \cite{stavrinos-ikeda 1999,Ikeda:2019ckp,Minas:2019urp}. 
Also, the causality problem and light cones for different types of Finsler 
spaces have been investigated 
in \cite{Caponio:2009er,Bernal:2020bul,Javaloyes:2018lex,Minguzzi:2014aua}. 
In this context the 
Raychaudhuri equations with locally anisotropic internal variables 
which are providing extra effective terms 
have been derived in \cite{Stavrinos:2012kv,Minguzzi:2015xka,Stavrinos:2016xyg,Triantafyllopoulos:2018bli,Relancio:2020rys}. 
Investigations of the extended Friedmann equations in Finsler spaces with extra 
internal degrees of freedom \cite{Stavrinos:2006rf} and dynamical analysis 
(critical points) \cite{Papagiannopoulos:2017whb,Papagiannopoulos:2020mmm}
provide a better understanding of the dynamical 
properties of the Finsler-Randers cosmological models.
To this end, articles in the framework of the weak field and pp-waves 
in Finsler spacetime can be found in 
\cite{Stavrinos:2012kv,Triantafyllopoulos:2018bli,Fuster:2015tua} and 
potentially they
can be used in order to test the performance of the  
Finslerian gravitational theory against current observations of 
gravitational waves.

It is well known in gravitation and cosmology that the Schwarzschild metric constitutes a fundamental ingredient of general relativity. This metric describes the most general spherical symmetric solution of the Einstein field equations in a region of spacetime where the energy-momentum tensor vanishes.

In the context of a Finsler/ generalized Finsler space, an extension of a locally anisotropic perturbation of a Finsler type Schwarzschild metric has been studied by different authors. In some of these works, possible observational predictions are given based on the direction-dependent structure of spacetime. We would like to point out that our study is realized in a Schwarzschild-Randers model which is different from the works of other researchers on the Finslerian extensions of a classic Schwarzschild metric \cite{Vacaru:2009sv,asanov1992s,rutz1993fs,Vacaru:2007tn,Silagadze:2010bi,Lammerzahl:2012kw,Vacaru:2009yi,Kinyanjui:2017xuk,perelman2018,Colladay:2019lig,Abraham:2020ytb}. In our approach, we use sufficiently generalized Einstein field equations on a Lorentz tangent bundle of a spacetime manifold.

In section \ref{sec: preliminaries} we introduce the basic framework and geometric structures of our model.

In section \ref{sec: Field equations} we study a Schwarzschild metric in a 
special Finsler-like spacetime of Randers type. 
This study provides a locally anisotropic perturbation of the classical Schwarzschild metric of the Riemannian structure in a natural way, induced by a covector field of the base manifold. In addition, the geometrical setting that we use, namely the framework of a Lorentz tangent bundle of a Riemannian spacetime, contains additional degrees of freedom compared to classic gravity. The generalized field equations which have been derived in \cite{Triantafyllopoulos:2020ogl} are applied on the perturbed metric of our Schwarzschild-Randers spacetime and are solved for the covector field.

In section \ref{sec: Paths}, we study particle paths for our generalized spacetime. We follow the approach in \cite{Triantafyllopoulos:2020ogl} which takes into account the effect of internal degrees of freedom on the point particle dynamics. We apply the solution of the covector derived in the previous sections and we obtain an explicit form for the path equations which is an extension of classical geodesics of general relativity. Finally, as an application, we solve the timelike paths numerically and compare them with the geodesics of general relativity.

\section{Preliminaries}\label{sec: preliminaries}
The natural background space for a locally anisotropic gravity is the tangent bundle of a differentiable Lorentzian spacetime manifold called a Lorentz Tangent Bundle (we will refer to it as $ TM $ hereafter) \cite{Miron:1994nvt,Vacaru:2005ht}. $TM$ is itself an 8-dimensional differentiable manifold, so we can define coordinate charts and tensors on it in the usual way. We briefly present the basics for this structure, for more details see \ref{sec:preliminaries}.

The Lorentz tangent bundle $TM$ is locally covered by a coordinate map $\{x^\mu,y^\alpha\}$ where the range of values for the indices of the $x$ variables is $\kappa,\lambda,\mu,\nu,\ldots = 0,\ldots,3$ and the range of values for the indices of the $y$ variables is  $\alpha,\beta,\ldots,\theta = 4,\ldots,7$.
An adapted basis on $TM$ is $ \left\{\delta_\mu = \frac{\partial}{\partial x^\mu} - N^\beta_\mu\frac{\partial}{\partial y^\beta}, \pdot{\alpha}= \frac{\partial}{\partial y^\alpha} \right\} $ and its dual basis is $ \left\{\de x^\mu,\delta y^\alpha = \de y^\alpha + N^\alpha_\nu \de x^\nu \right\} $.

The bundle $TM$ is equipped with a Sasaki-type metric $\Gd$:
\begin{equation}
	\mathcal{G} = g_{\mu\nu}(x,y)\,\mathrm{d}x^\mu \otimes \mathrm{d}x^\nu + v_{\alpha\beta}(x,y)\,\delta y^\alpha \otimes \delta y^\beta \label{bundle metric}
\end{equation}
where the metric of the horizontal space (h-space) $g_{\mu\nu}$ and the metric of the vertical space (v-space) $v_{\alpha\beta}$ are defined to be of Lorentzian signature $(-,+,+,+)$. In the rest of this work, the following homogeneity conditions will be assumed: $g_{\mu\nu}(x,ky) = g_{\mu\nu}(x,y)$, $v_{\alpha\beta}(x,ky) = v_{\alpha\beta}(x,y)$, $k>0$. These conditions are met when the following relations hold:
\begin{align}
	g_{\alpha\beta} = \pm\frac{1}{2}\pdder{F^2_g}{y^\alpha}{y^\beta} \label{Fg}\\
	v_{\alpha\beta} = \pm\frac{1}{2}\pdder{F^2_v}{y^\alpha}{y^\beta} \label{Fv}
\end{align}
where the functions $F_g$, $F_v$ satisfy the following conditions:
\begin{enumerate}
	\item $F_m$, $m=g,v$, is continuous on $TM$ and smooth on  $ \widetilde{TM}\equiv TM\setminus \{0\} $ i.e. the tangent bundle minus the null set $ \{(x,y)\in TM | F_m(x,y)=0\}$ \label{finsler field of definition}
	\item $ F_m $ is positively homogeneous of first degree on its second argument:
	\begin{equation}
		F_m(x^\mu,ky^\alpha) = kF_m(x^\mu,y^\alpha), \qquad k>0 \label{finsler homogeneity}
	\end{equation}
	\item The form 
	\begin{equation}
		f_{\alpha\beta}(x,y) = \dfrac{1}{2}\pdder{F_m^2}{y^\alpha}{y^\beta} \label{finsler metric} 
	\end{equation}
	defines a non-degenerate matrix: \label{finsler nondegeneracy}
	\begin{equation}
		\det\left[f_{\alpha\beta}\right] \neq 0 \label{finsler nondegenerate}
	\end{equation}
\end{enumerate}
with $ g_{\alpha\beta} = \tilde{\delta_\alpha^\mu}\tilde{\delta_\beta^\nu}g_{\mu\nu} $ and the sign in the rhs of \eqref{Fg}, \eqref{Fv} is chosen so that the resulting metric has the corrrect signature.

When the above conditions are met, our metric is called a (pseudo-)Finsler metric. Details about the connection, curvature and torsion structures can be found in \ref{sec:preliminaries}.

\section{Field equations}\label{sec: Field equations}
In this chapter we consider a general metric \eqref{bundle metric} on the tangent bundle of a spacetime manifold constituted from a horizontal and a vertical part. In this consideration, we set the horizontal part to be the classical Schwarzschild metric and the vertical part to be a Randers-type perturbation of the Schwarzschild metric, so the form of our metric will be a Schwarzschild-Randers metric. We also consider a similar case for a Schwarzschild-De Sitter-Randers metric. This is a different type of metric than the Finsler-Randers one which has been considered in the framework of cosmological study by different authors \cite{randers1940,Stavrinos:2005cgc,Stavrinos:2006rf,Basilakos:2013hua,Silva:2015ptj,Stavrinos:2016xyg,Chaubey:2018wph,Raushan:2020mkh,Papagiannopoulos:2020mmm}. In these cases, the horizontal part is a Friedmann-Robertson-Walker model. In both models the velocity and the vertical part play the role of intrinsic anisotropic perturbation of the traditional metrics.
We already know the form of the Schwarzschild metric, so we need to find an explicit form for the Randers-type perturbation which contains the additional information of local anisotropy.

\subsection{Field equations on the Lorentz tangent bundle}
In this paragraph, we present a set of field equations for the dynamic variables of our generalized framework. These equations are: 

\begin{align}
	& \overline R_{\mu\nu} - \frac{1}{2}({\overline R}+{S})\,{g_{\mu\nu}} \nonumber\\
	& \quad + \left(\delta^{(\lambda}_\nu\delta^{\kappa)}_\mu - g^{\kappa\lambda}g_{\mu\nu} \right)\left(\mathcal D_\kappa\mathcal T^\beta_{\lambda\beta} - \mathcal T^\gamma_{\kappa\gamma}\mathcal T^\beta_{\lambda\beta}\right)  = \kappa {T_{\mu\nu}} \label{feq1}\\
	& S_{\alpha\beta} - \frac{1}{2}({\overline R}+{S})\,{v_{\alpha\beta}} \nonumber\\
	& \quad + \left(v^{\gamma\delta}v_{\alpha\beta} - \delta^{(\gamma}_\alpha\delta^{\delta)}_\beta \right)\left(\mathcal D_\gamma C^\mu_{\mu\delta} - C^\nu_{\nu\gamma}C^\mu_{\mu\delta} \right) = \kappa {Y_{\alpha\beta}} \label{feq2}\\
	& g^{\mu[\kappa}\pdot{\alpha}L^{\nu]}_{\mu\nu} +  2 \mathcal T^\beta_{\mu\beta}g^{\mu[\kappa}C^{\lambda]}_{\lambda\alpha} = \frac{\kappa}{2}\mathcal Z^\kappa_\alpha \label{feq3}
\end{align}
with
\begin{align}
	T_{\mu\nu} &\equiv - \frac{2}{\sqrt{|\Gd|}}\frac{\Delta\left(\sqrt{|\Gd|}\,\mathcal{L}_M\right)}{\Delta g^{\mu\nu}} = - \frac{2}{\sqrt{-g}}\frac{\Delta\left(\sqrt{-g}\,\mathcal{L}_M\right)}{\Delta g^{\mu\nu}}\label{em1}\\
	Y_{\alpha\beta} &\equiv -\frac{2}{\sqrt{|\Gd|}}\frac{\Delta\left(\sqrt{|\Gd|}\,\mathcal{L}_M\right)}{\Delta v^{\alpha\beta}}  = -\frac{2}{\sqrt{-v}}\frac{\Delta\left(\sqrt{-v}\,\mathcal{L}_M\right)}{\Delta v^{\alpha\beta}}\label{em2}\\
	\mathcal Z^\kappa_\alpha &\equiv -\frac{2}{\sqrt{|\Gd|}}\frac{\Delta\left(\sqrt{|\Gd|}\,\mathcal{L}_M\right)}{\Delta N^\alpha_\kappa} = -2\frac{\Delta\mathcal{L}_M}{\Delta N^\alpha_\kappa}\label{em3}
\end{align}
where $\mathcal L_M$ is the Lagrangian of the matter fields, $\delta^\mu_\nu$ and $ \delta^\alpha_\beta$ are the Kronecker symbols, $|\Gd|$ is the absolute value of the determinant of the total metric \eqref{bundle metric}, and
\begin{equation}\label{torsion}
	\mathcal{T}_{\nu\beta}^{\alpha} = \pdot{\beta} N_{\nu}^{\alpha} - L_{\beta\nu}^{\alpha}
\end{equation}
are torsion components, where $L_{\beta\nu}^{\alpha}$ is defined in \eqref{metric d-connection 2}. From the form of \eqref{bundle metric} it follows that $\sqrt{|\Gd|} = \sqrt{-g}\sqrt{-v}$, with $g, v$ the determinants of the metrics $g_{\mu\nu}, v_{\alpha\beta}$ respectively. This relation was used in \eqref{em1}-\eqref{em3}.

Equations \eqref{feq1}-\eqref{feq3} are derived from an extension of the Hilbert-Einstein action on the 8 - dimensional Lorentz tangent bundle and constitute a generalization of the Einstein field equations of general relativity. They are appropriate for the study of locally anisotropic models of gravity with internal degrees of freedom. For details on their derivation from a Hilbert-like action on the Lorentz tangent bundle see \cite{Triantafyllopoulos:2020ogl}.

We will make some comments in order to give a physical interpretation in relation to the equations \eqref{em1}, \eqref{em2} and \eqref{em3}. Lorentz violations produce anisotropies in the space and the matter sector.
These act as a source of local anisotropy and can contribute to the torsion, connection, curvature components and to the energy-momentum tensors of the horizontal and vertical space $T_{\mu\nu}$ and $Y_{\alpha\beta}$. As a result, the curvatures $\overline R_{\mu\nu}$ and $R$ as well as the energy-momentum tensor $T_{\mu\nu}$ contain the additional information of local anisotropy of the metric and the matter fields. $S_{\alpha\beta}, S$ and $Y_{\alpha\beta}$, on the other hand, are objects with no equivalent in Riemannian gravity. They contain more information about local anisotropy which is produced from the metric $v_{\alpha\beta}$ which includes additional internal structure of spacetime. Finally, the nonlinear connection $N^\alpha_\mu$, a structure which induces an interaction between internal and external spaces \cite{Ikeda:2019ckp}, can also contribute to all the above-mentioned objects, while the energy-momentum tensor $ \mathcal Z^\kappa_\alpha $ shows the variation of $\mathcal L_M$ with respect to $N^\alpha_\mu$ and it 
reflects the dependence of matter fields on the nonlinear connection \cite{Triantafyllopoulos:2020ogl}. This is different from $T_{\mu\nu}$ and $Y_{\alpha\beta}$ which depend on just the external or internal structure respectively.

Notice that the Field equations \eqref{feq1}-\eqref{feq3} reduce 
to the usual Einstein field equations of general relativity (GR) in the limit:
\begin{align}
	g_{\mu\nu}(x,y) & \rightarrow g_{\mu\nu}(x)  \label{limit1}\\
	v_{\alpha\beta}(x,y) & \rightarrow g_{\alpha\beta}(x) = \tilde \delta^\mu_\alpha \tilde \delta ^\nu_\beta g_{\mu\nu}(x) \label{limit2} \\
	N^\alpha_\nu(x,y) & \rightarrow \frac{1}{2}y^\gamma g^{\alpha\beta}\partial_\nu g_{\beta\gamma}. \label{limit3}
\end{align}
We observe that when the metric $g_{\mu\nu}(x,y)$ is reduced to a Riemannian one according to \eqref{limit1} then the connection coefficients $L^\mu_{\kappa\lambda}(x,y)$ (eq. \eqref{metric d-connection 1}) reduce to the Christoffel symbols:
\begin{equation}\label{Christoffel2}
	\gamma^\mu_{\kappa\lambda}(x) = \frac{1}{2} g^{\mu\nu}\left( \partial_\kappa g_{\nu\lambda} + \partial_\lambda g_{\nu\kappa} - \partial_\nu g_{\kappa\lambda} \right)
\end{equation}
and the Cartan torsion tensor \eqref{metric d-connection 3} vanishes:
\begin{equation}\label{C=0}
	C^\mu_{\nu\alpha} = 0.
\end{equation} 
From \eqref{limit2} it becomes clear that the internal structure ($y$ variable) of spacetime doesn't provide additional information about the geometry, since the metric $v_{\alpha\beta}$ of the internal structure reduces to the metric $g_{\mu\nu}(x)$ of base spacetime ($x$ variable).

Moreover, we see that the Cartan-type \cite{Miron:1994nvt} 
nonlinear connection \eqref{limit3} ensures that  
$\mathcal T^\alpha_{\nu\alpha} = 0$. 
Indeed, from Eq. \eqref{torsion} and from the conditions (\ref{limit1}-\ref{limit3}) we get
\begin{align}
	\mathcal T^\alpha_{\nu\alpha} & = \pdot{\alpha}N^\alpha_\nu - \frac{1}{2} g^{\alpha\beta}\partial_\nu g_{\alpha\beta} \nonumber\\ & = \frac{1}{2} g^{\alpha\beta}\partial_\nu g_{\alpha\beta} - \frac{1}{2} g^{\alpha\beta}\partial_\nu g_{\alpha\beta} = 0.
\end{align}
Now, as long as conditions \eqref{limit1} and \eqref{limit2} are satisfied, the metric tensor \eqref{bundle metric} takes the Sasaki form \cite{Sas-58}
\begin{equation}
	\mathcal{G} = g_{\mu\nu}(x)\,\mathrm{d}x^\mu \otimes \mathrm{d}x^\nu + g_{\alpha\beta}(x)\,\delta y^\alpha \otimes \delta y^\beta \label{sasaki metric}
\end{equation}
Based on \eqref{sasaki metric}, it is straightforward to calculate the curvatures $\overline R_{\mu\nu}$ and $ \overline R$ from \eqref{Rmnline} and \eqref{Rline} respectively in the GR limit of the Lorentz tangent bundle gravity and we 
find that they reduce to the Ricci tensor $R_{\mu\nu}$ and Ricci scalar $R$ of general relativity for the metric $g_{\mu\nu}$. 
As expected, the curvatures $S_{\alpha\beta}$ 
and $S$ (Eqs.\eqref{d-ricci 4} and \eqref{hv ricci scalar}) both 
vanish in this limit.

From the above relations (\ref{limit1}-\ref{sasaki metric}) 
it follows that, in the GR limit, the corresponding 
field equations (\ref{feq1}-\ref{feq3}) boil down to 
\begin{align}
	R_{\mu\nu} - \frac{1}{2}R g_{\mu\nu} & = \kappa T_{\mu\nu} \label{feqlim1} \\
	-\frac{1}{2}R g_{\alpha\beta} & = \kappa Y_{\alpha\beta} \label{feqlim2} \\
	Z^\kappa_\alpha & = 0 \label{feqlim3}
\end{align}
where $R_{\mu\nu}$ and $R$ are the Ricci tensor and scalar 
of general relativity for the metric $g_{\mu\nu}$, as we mentioned above. 
Of course in Eq. \eqref{feqlim1} 
$G$ denotes the Newton's constant and 
$\kappa=\frac{8\pi G}{c^4}$, while the energy momentum 
tensor $T_{\mu\nu}$ is given by \eqref{em1}.
From the the trace of equation \eqref{feqlim1} we obtain $R=-\kappa T$, hence 
Eq.\eqref{feqlim2} gives
\begin{equation}
	Y_{\alpha\beta} = \frac{1}{2}Tg_{\alpha\beta}.
\end{equation}
This is the GR limit for the energy-momentum tensor $Y_{\alpha\beta}$.

Finally, from \eqref{em3}  and \eqref{feqlim3} 
we conclude that in the GR limit the matter fields 
have no direct dependence on the nonlinear connection.

\subsection{Schwarzschild-Randers spacetime}
As we mentioned above, the horizontal part $g_{\mu\nu}$ of the metric \eqref{bundle metric} will be taken to be the Schwarzschild metric so that
\begin{align}\label{Schwarzchild}
	&g_{\mu\nu}\de x^\mu  \de x^\nu \nonumber\\
	&\quad = -\left(1-\frac{R_s}{r}\right)dt^2 + \frac{dr^2}{1-\frac{R_s}{r}} + r^2 d\theta^2 + r^2 \sin^{2}\theta\, d\phi^2
\end{align}
where $R_s=2GM$ is the Schwarzschild radius (we have set the speed of light constant $c=1$). 

In the following, we assume a function $F_v$ of Randers type from which we will derive $v_{\alpha\beta}$ by using \eqref{Fv}:
\begin{equation}\label{RandersL}
	F_v = \sqrt{-g_{\alpha\beta}(x)y^\alpha y^\beta} + A_\gamma(x) y^\gamma
\end{equation}
where $g_{\alpha\beta}=g_{\mu\nu}\tilde\delta^{\mu}_{\alpha}\tilde\delta^{\nu}_{\beta}$ is the Schwarzschild metric and $A_{\gamma}(x)$ is a covector which will be determined by the equations. We focus on the timelike subspace of the internal $y-$space with respect to the Schwarzschild metric ($g_{\alpha\beta}(x)y^\alpha y^\beta <0$) hence the minus sign under the square root.
We take $A_{\gamma}(x)$ to be a weak term  ($|A_\gamma(x)|\ll 1$), hence
we neglect high order terms from the calculations.  
In addition, we consider as the appropriate form for the nonlinear connection the one given in \eqref{limit3} i.e. its GR limit value on the tangent bundle:
\begin{equation}\label{Nconnection}
	N^\alpha_\mu = \frac{1}{2}y^\beta g^{\alpha\gamma}\partial_\mu g_{\beta\gamma}
\end{equation}
This choice will give us a locally anisotropic gravitational model 
which deviates minimally from general relativity due to the extra Randers term $A_\gamma y^\gamma$ in \eqref{RandersL}.

We calculate the metric tensor $v_{\alpha\beta}$ of \eqref{RandersL} from \eqref{Fv}:
\begin{equation}\label{vRandersmetric}
	v_{\alpha\beta}= -\frac{1}{2}\frac{\partial^{2}F_v^2}{\partial y^{\alpha}\partial y^{\beta}}  
\end{equation}
and we find 
\begin{align}\label{vabfull}
	v_{\alpha\beta} = & \,g_{\alpha\beta}(x) + \frac{1}{a}(A_{\beta}g_{\alpha\gamma}y^\gamma + A_{\gamma}g_{\alpha\beta}y^\gamma + A_{\alpha}g_{\beta\gamma}y^\gamma) \nonumber\\
	& + \frac{1}{a^3}A_{\gamma}g_{\alpha\epsilon}g_{\beta\delta}y^\gamma y^\delta y^\epsilon , 
\end{align}
where we have set $a = \sqrt{-g_{\alpha\beta}y^{\alpha}y^{\beta}}$. From \eqref{vabfull} we see that the metric $v_{\alpha\beta}$ takes the form
\begin{equation}\label{vab}
	v_{\alpha\beta}(x,y) = g_{\alpha\beta}(x) + w_{\alpha\beta}(x,y)   
\end{equation}
where we have set
\begin{align}\label{wab}
	w_{\alpha\beta} = & \, \frac{1}{a}(A_{\beta}g_{\alpha\gamma}y^\gamma + A_{\gamma}g_{\alpha\beta}y^\gamma + A_{\alpha}g_{\beta\gamma}y^\gamma) \nonumber\\
	& + \frac{1}{a^3}A_{\gamma}g_{\alpha\epsilon}g_{\beta\delta}y^\gamma y^\delta y^\epsilon
\end{align}
Its inverse is $v^{\beta\gamma} = g^{\beta\gamma} - w^{\beta\gamma}$ so that $v_{\alpha\beta} v^{\beta\gamma} = g_{\alpha\beta}g^{\beta\gamma} = \delta_\alpha^\gamma$ to first order in $w_{\alpha\beta}$. The total metric over the tangent bundle is then written as
\begin{equation}\label{bundle metric 3}
	\mathcal{G} = g_{\mu\nu}(x)\, \de x^{\mu}\otimes \de x^{\nu} + \left[g_{\alpha\beta}(x) + w_{\alpha\beta}(x,y)\right]\, \delta y^{\alpha}\otimes \delta y^{\beta}
\end{equation}
We remark that, as we can see from \eqref{vab}, the metric $v_{\alpha\beta}(x,y)$ is a Finslerian perturbation of the Riemannian metric $g_{\alpha\beta}(x)$. 
We observe that if we let $w_{\alpha\beta}\rightarrow 0$ then the field equations \eqref{feq1}-\eqref{feq3} reduce to the Einstein field equations of general relativity.

Next we will calculate the terms for \eqref{feq1} and \eqref{feq2}. From the definitions \eqref{Rmnline} and \eqref{Rline} we see that when $g_{\mu\nu}$ has no explicit dependence on $y$ then $ \overline R_{\mu\nu} $ and $ \overline R $ reduce to the classical Ricci tensor and scalar of general relativity. Additionally, since $g_{\mu\nu}(x)$ is the Schwarzschild metric, both $\overline R_{\mu\nu}$ and $\overline R $ are zero. Here we have assumed vacuum solutions, so the energy momentum tensors are zero and the equations \eqref{feq1} and \eqref{feq2} become
\begin{equation}\label{feq1v2}
	-\frac{1}{2}S g_{\mu\nu} +  \left(\delta^{(\lambda}_\nu\delta^{\kappa)}_\mu - g^{\kappa\lambda}g_{\mu\nu}\right) \left(\mathcal D_\kappa\mathcal T^\beta_{\lambda\beta} - \mathcal T^\gamma_{\kappa\gamma}\mathcal T^\beta_{\lambda\beta}\right) = 0
\end{equation}
\begin{align}\label{feq2v2}
	&S_{\alpha\beta} - \frac{1}{2}S\,v_{\alpha\beta} \nonumber\\
	& \quad + \left(v^{\gamma\delta}v_{\alpha\beta} - \delta^{(\gamma}_\alpha\delta^{\delta)}_\beta \right)\left(\mathcal D_\gamma C^\mu_{\mu\delta} - C^\nu_{\nu\gamma}C^\mu_{\mu\delta} \right) = 0 
\end{align}
Field equation \eqref{feq3} gives us no additional information since all three terms vanish identically in our case. We can simplify equation \eqref{feq2v2} by calculating $C_{\mu\delta}^{\mu}$ from \eqref{metric d-connection 3} and we find that it is zero since the metric $g_{\mu\nu}$ depends only on $x$. Then by taking the trace of the remaining terms in \eqref{feq2v2} we can show that $S_{\alpha\beta}$ and $S$ are also zero, so the field equation \eqref{feq1v2} becomes
\begin{equation}\label{hfeq}
	\left(\delta^{(\lambda}_\nu\delta^{\kappa)}_\mu - g^{\kappa\lambda}g_{\mu\nu}\right)\left(\mathcal D_\kappa\mathcal T^\beta_{\lambda\beta} - \mathcal T^\gamma_{\kappa\gamma}\mathcal T^\beta_{\lambda\beta}\right) = 0.
\end{equation}
We substitute \eqref{metric d-connection 2} and \eqref{Nconnection} in \eqref{torsion} and after some calculations we get
\begin{equation}
	\mathcal T^\alpha_{\nu\alpha} = -\frac{1}{2}\delta_\nu w\label{tors}
\end{equation}
with $w=g_{\alpha\beta} w^{\alpha\beta}$. The above relation \eqref{tors} shows us that the torsion is of first order on $w_{\alpha\beta}$ so the terms  $\mathcal T^\gamma_{\kappa\gamma}\mathcal T^\beta_{\lambda\beta}$ from \eqref{hfeq} are omitted. Then by taking the trace of the remaining terms in \eqref{hfeq} we have the equation that follows:
\begin{equation}\label{DTtrace}
	g^{\mu\nu}\mathcal D_\mu\mathcal T^\alpha_{\nu\alpha} = 0
\end{equation}
Substituting the latter equation to \eqref{hfeq} we find
\begin{equation}\label{DmTn}
	\mathcal D_{(\mu}\mathcal T^{\alpha}_{\nu)\alpha} = 0
\end{equation}
By the definition of the covariant derivative in \eqref{vector h-covariant}, equation \eqref{DmTn} becomes
\begin{equation}\label{covT}
	\delta_{(\mu}\mathcal T^\alpha_{\nu)\alpha}-L_{\mu\nu}^{\kappa}\mathcal T^\alpha_{\kappa\alpha} = 0
\end{equation}
Using equations \eqref{wab} and \eqref{tors} we get $\mathcal T^\alpha_{\nu\alpha} $ in terms of $A_\beta$:
\begin{equation}\label{T(A)}
	\mathcal T^\alpha_{\nu\alpha} = -\frac{5}{2}\delta_\nu \left( A_\beta \frac{y^\beta}{a} \right).
\end{equation}
It is straightforward to show that
\begin{equation}\label{dy/a}
	\delta_\mu\left( \frac{y^\alpha}{a}\right) = -E^\gamma_{\beta\mu} \frac{y^\beta}{a}
\end{equation}
with
\begin{equation}\label{E}
	E^\gamma_{\beta\mu}(x) \equiv \frac{1}{2} g^{\alpha\gamma} \partial_\mu g_{\alpha\beta}
\end{equation}
Using \eqref{T(A)} and \eqref{dy/a} we get
\begin{equation}\label{T(KA)}
	\mathcal T^\alpha_{\nu\alpha} = -\frac{5}{2}K_{\nu\gamma} \frac{y^\gamma}{a}
\end{equation}
with
\begin{equation}\label{K}
	K_{\nu\gamma}(x) \equiv \partial_\nu A_\gamma - A_\beta E^\beta_{\gamma\nu} .
\end{equation}
from relation (\ref{T(KA)}) we can calculate \eqref{covT}:
\begin{equation}\label{Kdiff}
	\left( \partial_{(\mu}K_{\nu)\gamma} - E^\beta_{\gamma(\mu}K_{\nu)\beta} - L^\lambda_{\mu\nu}K_{\lambda\gamma} \right) \frac{y^\gamma}{a} = 0 .
\end{equation}
Relation \eqref{Kdiff} must hold for every $y$. Since the expression in parentheses does not depend on $y$, we conclude that it must identically vanish:
\begin{equation}\label{Kdiff0}
	\partial_{(\mu}K_{\nu)\gamma} - E^\beta_{\gamma(\mu}K_{\nu)\beta} - L^\lambda_{\mu\nu}K_{\lambda\gamma} = 0
\end{equation}
\\

{\bf Remark:} By comparing \eqref{metric d-connection 2} and \eqref{E} we get a relation of the form $L^\alpha_{\beta\mu} = E^\alpha_{\beta\mu} + \mathcal O(A)$. If we use this in \eqref{K} and \eqref{Kdiff0} we get the equation $\mathcal D_{(\mu}\mathcal D_{\nu)}A_\gamma = 0$, which is equivalent to the system of equations \eqref{K} and \eqref{Kdiff0}. Contracting this with $g^{\mu\nu}$ gives
\begin{equation}\label{boxA}
	\square A_\gamma = 0
\end{equation}
with $\square \equiv g^{\mu\nu}\mathcal D_\mu \mathcal D_\nu$.
\\

In order to fully determine the metric \eqref{bundle metric 3} for the respective subspace of the internal space, we need $g_{\mu\nu}(x)$ and $A_\gamma(x)$. The first is already defined in \eqref{Schwarzchild}, so we need to solve (\ref{Kdiff0}) for $A_\gamma(x)$ to get a full expression for the metric in our space.
If we use the definitions \eqref{E} and \eqref{K} on relation \eqref{Kdiff0}, we get the equation that we need to solve for $A(x)$:
\begin{align}
	&\partial_\mu \partial_\nu A_\gamma - \frac{1}{2}g^{\beta\delta}\partial_\nu g_{\delta\gamma}\partial_\mu A_\beta - \frac{1}{2}g^{\beta\delta}\partial_\mu g_{\delta\gamma}\partial_\nu A_\beta \nonumber \\
	& +\frac{1}{4} A_\beta \Bigg( \frac{1}{2} g^{\beta\epsilon} g^{\delta\zeta} \partial_\nu g_{\epsilon\delta} \partial_\mu g_{\gamma\zeta} + \frac{1}{2} g^{\beta\epsilon} g^{\delta\zeta} \partial_\mu g_{\epsilon\delta}  \partial_\nu g_{\gamma\zeta} \nonumber\\
	& - \partial_\mu g^{\beta\delta} 
	\partial_\nu g_{\gamma\delta} - \partial_\nu g^{\beta\delta} 
	\partial_\mu g_{\gamma\delta} - 2g^{\beta\delta}\partial_\mu\partial_\nu g_{\gamma\delta} \Bigg) \nonumber \\
	& -\frac{1}{2}g^{\kappa\lambda} \left( \partial_\mu g_{\kappa\nu} + \partial_\nu g_{\kappa\mu} - \partial_\kappa g_{\mu\nu} \right) \nonumber\\
	&\quad\times\left( \partial_\lambda A_\gamma - \frac{1}{2} A_\beta g^{\beta\delta} \partial_\lambda g_{\gamma\delta} \right) = 0 \label{fullexp},
\end{align}
where $g_{\mu\nu}(x)$ is the Schwarzschild metric \eqref{Schwarzchild}. Once we get $A(x)$ from \eqref{fullexp}, we can calculate $w_{\alpha\beta}(x,y)$ from \eqref{wab} and then use the result to calculate the full metric \eqref{bundle metric 3}.

A similar analysis holds for the spatial subspace of the internal space. In that case, instead of \eqref{RandersL} and \eqref{vRandersmetric} we have
\begin{equation}
	F_v = \sqrt{g_{\alpha\beta}y^\alpha y^\beta} + A_\gamma y^\gamma
\end{equation}
and
\begin{align}\label{spacelike_v}
	v_{\alpha\beta} = & \, \frac{1}{2}\frac{\partial^{2}F_v^2}{\partial y^{\alpha}\partial y^{\beta}} \nonumber\\
	= & \, g_{\alpha\beta}(x) + \frac{1}{a}(A_{\beta}g_{\alpha\gamma}y^\gamma + A_{\gamma}g_{\alpha\beta}y^\gamma + A_{\alpha}g_{\beta\gamma}y^\gamma) \nonumber\\
	& - \frac{1}{a^3}A_{\gamma}g_{\alpha\epsilon}g_{\beta\delta}y^\gamma y^\delta y^\epsilon ,
\end{align}
where $a = \sqrt{g_{\mu\nu}y^\mu y^\nu}$ for the spacelike sector of $g_{\mu\nu}$ taking into consideration the signature of the metric. Following the same steps as above, we reach the same equation for $A_\gamma$, i.e. eq.\eqref{fullexp}. Therefore, solving this equation will give us the metric for both the timelike and spacelike (with respect to $g_{\alpha\beta}$) sub-spaces of the internal space.

We will solve \eqref{fullexp} analytically with separation of variables, see \ref{Acalculation} for more details. After calculations we find the solution
\begin{equation}\label{Asolution}
	A_\gamma(x) = \left[ \tilde A_4 \left|1-\frac{R_S}{r} \right|^{1/2}, 0, 0, 0 \right]
\end{equation}
with $\tilde A_4$ a constant. This is a timelike covector since $g^{\alpha\beta}A_\alpha A_\beta = -(\tilde A_4)^2 < 0$.
It is interesting to mention that the horizon of the Schwarzschild - Randers 
metric is correlated with that of Schwarzschild.
Practically, the quantity $A_\gamma$ can be seen as a distortion factor which
quantifies the deviation from the pure Schwarzschild solution. 
Obviously, the solution \eqref{Asolution}
on small spherical scales ($r\sim R_{s}$) tends to zero. On the other hand, 
for $r \gg R_{s}$ the Schwarzschild - Randers metric tends asymptotically to Minkowski. Finally, we see that this metric has a singularity in $r=0$ similarly with the classic Schwarzschild one. The Schwarzschild - Randers model will be further studied for intrinsic singularities and horizons in a future research.

\subsection{Schwarzschild-De Sitter-Randers spacetime}

We will follow the same procedure as in the previous paragraph but for a Schwarzschild-Randers spacetime with a cosmological horizon, namely a Schwarzschild-De Sitter-Randers spacetime. In this scenario, we take the horizontal part of the metric \eqref{bundle metric} to be:
\begin{align}\label{Schwarzchild-desitter}
	g_{\mu\nu}\de x^\mu  \de x^\nu = & -\left(1-\frac{R_s}{r} - \frac{\Lambda}{3} r^2\right)dt^2 \nonumber\\
	&  + \frac{dr^2}{1-\frac{R_s}{r} - \frac{\Lambda}{3} r^2} + r^2 d\theta^2 + r^2 \sin^{2}\theta \, d\phi^2
\end{align}
while, as before, the metric tensor $v_{\alpha\beta} $ will be derived from the Lagrangian \eqref{RandersL} and relation \eqref{vRandersmetric},
where $g_{\alpha\beta} = \tilde \delta^\mu_\alpha \tilde \delta^\nu_\beta$ is now given by \eqref{Schwarzchild-desitter}. The latter is a static spherically symmetric vacuum solution for the classical Einstein field equations with a cosmological constant $\Lambda$:
\begin{equation}\label{EFE_L}
	R_{\mu\nu} - \frac{1}{2}g_{\mu\nu}R + g_{\mu\nu}\Lambda = 0
\end{equation}
with $R_{\mu\nu}$ and $R$ the Ricci tensor and scalar of general relativity. In accordance, we introduce a cosmological constant term to the field equations \eqref{feq1} in vacuum:
\begin{align}
	& {\overline R_{\mu\nu}} - \frac{1}{2}({\overline R} + {S})\,{g_{\mu\nu}} \nonumber\\
	& + \left(\delta^{(\lambda}_\nu\delta^{\kappa)}_\mu - g^{\kappa\lambda}g_{\mu\nu} \right)\left(\mathcal D_\kappa\mathcal T^\beta_{\lambda\beta} - \mathcal T^\gamma_{\kappa\gamma}\mathcal T^\beta_{\lambda\beta}\right) + g_{\mu\nu}\Lambda = 0 \label{feq1ds}
\end{align}
The tensors $\overline R_{\mu\nu}$ and $\overline R$ reduce to the standard Ricci tensor and scalar of general relativity for the metric \eqref{Schwarzchild-desitter} since the latter has no direct dependence on $y$. Additionally, from \eqref{feq2} in vacuum we get $S=0$ the same way as in the previous paragraph. Therefore, using \eqref{EFE_L} in equation \eqref{feq1ds} we get relation \eqref{hfeq} again. It is obvious that the procedure is the same as before and only the explicit form of $g_{\mu\nu}(x)$ changes. As such, we reach the same equation for $A_\gamma$, namely relation \eqref{fullexp} with $g_{\mu\nu}$ given by \eqref{Schwarzchild-desitter}. Again, by separation of variables, one finds (see \ref{Acalculation}):
\begin{equation}\label{Asolution2}
	A_\gamma(x) = \left[ \tilde A_4 \left|1-\frac{R_S}{r} - \frac{\Lambda}{3}r^2 \right|^{1/2}, 0, 0, 0 \right]
\end{equation}

\section{Paths in the Schwarzschild-Randers spacetime}\label{sec: Paths}
Now that we have $A_\gamma$ and hence the full metric, we can study particle trajectories in $TM$. A  Lagrangian for point particles in the total space of the Lorentz tangent bundle has been proposed in \cite{Triantafyllopoulos:2020ogl}:
\begin{equation}\label{full lagrangian}
	L(x,\dot x,y) = \left(a g_{\mu\nu}\dot x^\mu \dot x^\nu + b \tilde \delta^\alpha_\mu v_{\alpha\beta}\dot x^\mu y^\beta + c v_{\alpha\beta}y^\alpha y^\beta \right)^{1/2}
\end{equation}
with $a, b, c$ constants. The associated equations of motion are
\begin{equation}\label{genpath}
	\left(g_{\kappa\nu} + z \tilde\delta^\alpha_\kappa \delta^\beta_\nu v_{\alpha\beta}\right)\ddot x^\kappa + \left(\gamma_{\nu\kappa\lambda} + z\sigma_{\nu\kappa\lambda} \right) \dot x^\kappa \dot x^\lambda = 0
\end{equation}
and
\begin{equation}\label{y=xdot}
	y^\alpha = \tilde\delta^\alpha_\mu \dot x^\mu 
\end{equation}
with 
\begin{equation}
	\sigma_{\nu\kappa\lambda} = \frac{1}{2} \left(\tilde\delta^\alpha_\nu \tilde\delta^\beta_\lambda \partial_\kappa v_{\alpha\beta} + \tilde\delta^\alpha_\nu \tilde\delta^\beta_\kappa \partial_\lambda v_{\alpha\beta} - \tilde\delta^\alpha_\kappa \tilde\delta^\beta_\lambda \partial_\nu v_{\alpha\beta} \right)
\end{equation}
and $z = -b^2/4ac$ is a constant. The Christoffel symbols of the first kind for the metric $g_{\kappa\nu}(x)$ are
\begin{equation}\label{Christoffel1}
	\gamma_{\nu\kappa\lambda} = \frac{1}{2} \left( \partial_\kappa g_{\nu\lambda} + \partial_\lambda g_{\nu\kappa} - \partial_\nu g_{\kappa\lambda} \right)
\end{equation}

The term $ \tilde\delta^\alpha_\kappa \tilde\delta^\beta_\nu v_{\alpha\beta} $ is the metric of the v-space lowered down to the h-space via the generalized Kronecker symbols which are defined as $\tilde{\delta_\alpha^\mu} = \tilde\delta^\alpha_\mu = 1$ for $a=\mu+4$ and equal to zero otherwise\footnote{In general, $\tilde \delta^\alpha_\mu$ and $\tilde \delta_\beta^\nu$ can be used to lift an object from the horizontal to the vertical subspace of $TTM$ or lower down one from the vertical to the horizontal subspace. This allows us to perform algebraic operations between components of tensors belonging to different sub-spaces of $TTM$.}. We will write for convenience $ \tilde\delta^\alpha_\kappa \tilde\delta^\beta_\nu v_{\alpha\beta} = v_{\kappa\nu}$ and similarly $ \tilde\delta^\alpha_\kappa \tilde\delta^\beta_\nu w_{\alpha\beta} = w_{\kappa\nu}$.

We define $\overline g_{\kappa\nu} = g_{\kappa\nu} + zv_{\kappa\nu}$ and we observe that its inverse is $\overline g^{\mu\nu} = (1+z)^{-2}\left( g^{\mu\nu} + zv^{\mu\nu} \right)$ in the sense that $ \overline g_{\kappa\nu} \overline g^{\mu\nu} = \delta^\mu_\kappa $ to first order in $w_{\mu\nu}$. Contracting \eqref{genpath} with $ \overline g^{\mu\nu} $ gives
\begin{align}
	\overline g^{\mu\nu} \overline g_{\kappa\nu} \ddot x^\kappa + & \left( \overline g^{\mu\nu} \gamma_{\nu\kappa\lambda} + z \overline g^{\mu\nu} \sigma_{\nu\kappa\lambda} \right) \dot x^\kappa \dot x^\lambda = 0\nonumber \\
	\Leftrightarrow \ddot x^\mu + & \, (1+z)^{-2} \big[\gamma^\mu_{\kappa\lambda} + z \sigma^\mu_{\kappa\lambda} \nonumber\\
	+ & \, z v^{\mu\nu} \left( \gamma_{\nu\kappa\lambda} + z \sigma_{\nu\kappa\lambda} \right) \big] \dot x^\kappa \dot x^\lambda = 0 \label{genpath1}
\end{align}
where  $ \gamma^\mu_{\kappa\lambda} = g^{\mu\nu} \gamma_{\nu\kappa\lambda} $ and $ \sigma^\mu_{\kappa\lambda} = g^{\mu\nu} \sigma_{\nu\kappa\lambda} $. After some straightforward calculations, eq. \eqref{genpath1} gives
\begin{equation}\label{genpath2}
	\ddot x^\mu + \gamma^\mu_{\kappa\lambda} \dot x^\kappa \dot x^\lambda = - \frac{z}{1+z} \left( \tilde \sigma^\mu_{\kappa\lambda} - w^{\mu\nu} \gamma_{\nu\kappa\lambda} \right) \dot x^\kappa \dot x^\lambda ,
\end{equation}
with
\begin{equation}\label{tsigma}
	\tilde \sigma^\mu_{\kappa\lambda} \equiv \frac{1}{2} g^{\mu\nu} \left(\partial_\kappa w_{\nu\lambda} + \partial_\lambda w_{\nu\kappa} - \partial_\nu w_{\kappa\lambda} \right)
\end{equation}

The horizontal part of the tangent vector on the paths is $\dot x^\mu = dx^\mu/ds$ with $s$ an affine parameter along the path defined as \cite{Triantafyllopoulos:2020ogl}:
\begin{equation}\label{saffine}
	s = s_0 + \int_{\lambda_0}^{\lambda_1} \sqrt{\pm\overline g_{\mu\nu}(x,y) \frac{dx^\mu}{d\lambda} \frac{dx^\nu}{d\lambda}} \,d\lambda
\end{equation}
with $s_0$, $\lambda_0$ and $\lambda_1$ constants and $\lambda$ is an arbitrary parameter of the path. The sign of $ \overline g_{\mu\nu}(x,y) $ is determined by the tangent vector of the path, specifically if $d x^\nu/d\lambda$ is timelike with respect to $ \overline g_{\mu\nu}(x,y) $ ($ \overline g_{\mu\nu}(x,y) \frac{dx^\mu}{d\lambda} \frac{dx^\nu}{d\lambda} < 0 $) then we get $``-"$, likewise for a spacelike tangent vector with respect to $ \overline g_{\mu\nu}(x,y) $ ($ \overline g_{\mu\nu}(x,y) \frac{dx^\mu}{d\lambda} \frac{dx^\nu}{d\lambda} > 0 $) we get $``+"$.

The paths \eqref{genpath2} will play the role for our model that the geodesics play for general relativity. As is the case for the latter, we need a classification of path segments with respect to their character i.e. timelike, null and spacelike. We define:
\begin{itemize}
	\item Timelike segment: $g_{\mu\nu}(x) \dot x^\mu \dot x^\nu < 0 $ at every point
	\item Null segment: $g_{\mu\nu}(x) \dot x^\mu \dot x^\nu = 0 $ at every point
	\item Spacelike segment: $g_{\mu\nu}(x) \dot x^\mu \dot x^\nu > 0 $ at every point
\end{itemize}
Therefore, the character of the path is determined by the metric tensor $g_{\mu\nu}(x)$ of the horizontal subspace.
We define the proper time $\tau$ as
\begin{equation}\label{propertime}
	\tau = \tau_0 + \int_{\lambda_0}^{\lambda_1} \sqrt{-g_{\mu\nu}(x) \frac{dx^\mu}{d\lambda} \frac{dx^\nu}{d\lambda}} \,d\lambda ,
\end{equation}
where $\tau_0$ is constant. By comparing relations \eqref{saffine} and \eqref{propertime} we see that the parameter $s$ on the paths \eqref{genpath2} cannot be written as an affine transformation of the proper time in general.

We remark that equation \eqref{genpath2} reduces to the classic geodesics equation of general relativity when the perturbation $w_{\alpha\beta}$ goes to zero, as it should.

\subsection{Timelike paths}
To begin, we rewrite the perturbation \eqref{wab} as
\begin{equation}\label{wnl_time}
	w_{\nu\lambda} = g_{\lambda\rho} A_\nu u^\rho + g_{\nu\rho} A_\lambda u^\rho + \left( g_{\nu\lambda} + g_{\nu\sigma} g_{\lambda\tau} u^\sigma u^\tau \right) A_\rho u^\rho 
\end{equation}
with
\begin{equation}
	u^\nu \equiv \frac{y^\nu}{a}
\end{equation}
where $a = \sqrt{-g_{\mu\nu}y^\mu y^\nu}$ and we have lowered down $A_\gamma$ and $y^\gamma$ using the generalized Kronecker deltas. It is straightforward to show
\begin{equation}\label{pdu}
	\partial_\mu u^\nu = -\frac{1}{2} \partial_\mu g_{\kappa\lambda} u^\kappa u^\lambda u^\nu
\end{equation}
Using \eqref{tsigma}, \eqref{wnl_time} and \eqref{pdu} we calculate
\begin{align}
	\tilde\sigma^\mu_{\kappa\lambda}y^\kappa y^\lambda = & \, a^2 g^{\mu\nu} \Big\{ \Big(  g_{\lambda\sigma}\partial_\kappa A_\nu + 2g_{\nu\sigma} \partial_\kappa A_\lambda - \frac{3}{2}g_{\lambda\sigma} \partial_\nu A_\kappa \nonumber\\
	+ & \, 2A_\sigma \partial_\kappa g_{\nu\lambda} + A_\nu \partial_\kappa g_{\lambda\sigma} - \frac{3}{2} A_\sigma \partial_\nu g_{\kappa\lambda} \Big) u^\sigma u^\kappa u^\lambda \nonumber\\
	+ & \, \frac{1}{2} \Big( - A_\nu g_{\lambda\rho} \partial_\kappa g_{\sigma\tau} + A_\lambda g_{\nu\rho} \partial_\kappa g_{\sigma\tau} \nonumber\\
	+ & \, 2 A_\rho g_{\lambda\tau} \partial_\kappa g_{\nu\sigma} + 2g_{\nu\sigma}g_{\lambda\tau} \partial_\kappa A_\rho - A_\kappa g_{\lambda\rho} \partial_\nu g_{\sigma\tau} \nonumber\\
	- & \, g_{\kappa\sigma} g_{\lambda\tau} \partial_\nu A_\rho \Big) u^\sigma u^\tau u^\rho u^\kappa u^\lambda \nonumber \\
	- & \frac{3}{4} A_\rho \Big( 2g_{\nu\sigma} g_{\kappa\tau} \partial_\lambda g_{\xi\pi} \nonumber\\
	- & \, g_{\kappa\sigma} g_{\lambda\tau} \partial_\nu g_{\xi\pi} \Big)  u^\sigma u^\tau u^\rho u^\kappa u^\lambda u^\xi u^\pi \Big\}
\end{align}
Now, if we take into account that $g_{\mu\nu}u^\mu u^\nu = -1$, the above relation gives
\begin{align}\label{tsigmayy}
	\tilde\sigma^\mu_{\kappa\lambda}y^\kappa y^\lambda = & \, a^2 \Big\{ g^{\mu\nu} \left( \partial_\nu A_\kappa - \partial_\kappa A_\nu \right) u^\kappa \nonumber\\
	& + g^{\mu\nu} \Big( g_{\nu\sigma} \partial_\kappa A_\lambda + A_\sigma \partial_\kappa g_{\nu\lambda} + \frac{3}{2}A_\nu \partial_\kappa g_{\lambda\sigma} \nonumber\\
	& - \frac{1}{4}A_\kappa \partial_\nu g_{\sigma\lambda} \Big) u^\sigma u^\kappa u^\lambda \nonumber \\
	& + \, A_\lambda \partial_\kappa g_{\sigma\tau} u^\sigma u^\tau u^\mu u^\kappa u^\lambda \Big\} .
\end{align}
Substituting the relations \eqref{y=xdot}, \eqref{Christoffel1}, \eqref{wnl_time} and \eqref{tsigmayy} into \eqref{genpath2} we get
\begin{align}\label{timelike_path}
	& \ddot x^\mu + \gamma^\mu_{\kappa\lambda} \dot x^\kappa \dot x^\lambda \nonumber \\
	& = -\frac{z}{1+z}  \Big\{ a g^{\mu\nu} \left(\partial_\nu A_\kappa - \partial_\kappa A_\nu \right) \dot x^\kappa \nonumber\\
	& + \frac{1}{a} \Big[A^\nu \left(\partial_\kappa g_{\nu\lambda} - \frac{1}{2}\partial_\nu g_{\kappa\lambda} \right) + \partial_\kappa A_\lambda \Big] \dot x^\mu \dot x^\kappa \dot x^\lambda \nonumber\\
	& + \frac{1}{a} \Big( \frac{1}{4}g^{\mu\nu}\! A_\kappa \partial_\nu g_{\sigma\lambda} + g^{\mu\nu}\! A_\kappa \partial_\lambda g_{\sigma\nu} + A^\mu \partial_\kappa g_{\lambda\sigma} \Big) \dot x^\sigma \dot x^\kappa \dot x^\lambda \nonumber \\
	& + \frac{1}{2a^3} A_\lambda \partial_\kappa g_{\sigma\tau} \dot x^\sigma \dot x^\tau \dot x^\mu \dot x^\kappa \dot x^\lambda \Big\}
\end{align}
This is the generalized path equation for the timelike sector of the metric $g_{\mu\nu}(x)$.

{\bf Remark:} If we set $a=1$ at some fixed point then \eqref{timelike_path} can be written as
\begin{align}\label{timelike_path_lorentz}
	& \ddot x^\mu + \gamma^\mu_{\kappa\lambda} \dot x^\kappa \dot x^\lambda - \frac{e}{m}\tensor{F}{^\mu_\kappa}\dot x^\kappa \nonumber \\
	& = \frac{e}{m} \Big\{ \Big[A^\nu \left(\partial_\kappa g_{\nu\lambda} - \frac{1}{2}\partial_\nu g_{\kappa\lambda} \right) + \partial_\kappa A_\lambda \Big] \dot x^\mu \dot x^\kappa \dot x^\lambda \nonumber\\
	& + \Big( \frac{1}{4}g^{\mu\nu}\! A_\kappa \partial_\nu g_{\sigma\lambda} + g^{\mu\nu}\! A_\kappa \partial_\lambda g_{\sigma\nu} + A^\mu \partial_\kappa g_{\lambda\sigma} \Big) \dot x^\sigma \dot x^\kappa \dot x^\lambda \nonumber \\
	& + A_\lambda \partial_\kappa g_{\sigma\tau} \dot x^\sigma \dot x^\tau \dot x^\mu \dot x^\kappa \dot x^\lambda \Big\}
\end{align}
with $F_{\kappa\nu} = \partial_\nu A_\kappa - \partial_\kappa A_\nu$ the field strength tensor of $A_\nu$ and we have set $\frac{z}{1+z} := -\frac{e}{m}$ where $e$ the electric charge and $m$ the mass of the particle. If we ignore the r.h.s of the above equation then \eqref{timelike_path_lorentz} will have the same form as the equation of a charged particle subject to the Lorentz force with an electromagnetic vector potential $A_\gamma$ in the Riemannian setting. A similar equation which is derived from a Finsler-Randers Lagrangian and contains a Lorentz force term has been studied in \cite{Stavrinos:2016xyg}. However, in our more generalized setting we also get the r.h.s. perturbation term which depends on $A_\nu$ and its first derivatives. Therefore, a possible relation between our Schwarzschild-Randers metric and the Lorentz force requires further investigation and goes beyond the scope of this work.

Now, it is known that we can always approach a timelike path (geodesic) with a proper time parameter broken null path with the same endpoints \cite{carroll2004}. In this approximation it is considered that the number of null path segments with infinitesimal distance between two neighboring points increases following the timelike path. Therefore, the final null path of zero length (with respect to $g_{\mu\nu}$) approaches the timelike path \eqref{timelike_path}, however the parameter along them is replaced by an appropriate affine one.

Substituting to \eqref{timelike_path} the solution \eqref{Asolution} we get the explicit form of the timelike paths components for $r>R_s$:
\begin{align}
	&\ddot t + \frac{1-f}{rf}\dot r \dot t \nonumber\\
	& = -\frac{z}{1+z} \tilde A_4 \Big\{ \Big(\frac{1}{2}af^{-3/2} \dot r + \frac{2}{a}f^{-1/2} \dot r \dot t^2 \nonumber\\
	& + \frac{1}{a}f^{-5/2}\dot r^3 \Big)\frac{1-f}{r} \nonumber\\
	& -\frac{2r}{a}f^{-1/2} \big( \dot r \dot \theta^2 + \sin^2\theta \, \dot r \dot \phi^2 + \frac{r}{2}\sin 2\theta \, \dot \theta \dot \phi^2 \big) \nonumber\\
	& + \, \dot t \Big[ \frac{1}{a}\Big( -f^{-3/2} + \frac{1}{2}f^{-1/2} \Big) \dot t \dot r \frac{1-f}{r} \nonumber\\
	& + \, \frac{1}{2a^3} \Big( - \big\{ f^{1/2} \dot t^3 \dot r + f^{-3/2} \dot t \dot r^3 \big\} \frac{1-f}{r} \nonumber\\
	& + 2f^{1/2}r \big\{ \dot t \dot r \dot \theta^2 + \sin^2 \theta \, \dot t \dot r \dot \phi^2 + \frac{r}{2} \sin 2\theta \, \dot t \dot \theta \dot \phi^2 \big\} \Big) \Big]\Big\} \label{path0}
\end{align}
\begin{align}
	&\ddot r + \frac{f(1-f)}{2r} \dot t^2 - \frac{1-f}{2rf} \dot r^2 - rf \big( \dot \theta^2 + \sin^2 \theta \dot  \phi^2 \big) \nonumber\\
	& = -\frac{z}{1+z} \tilde A_4 \Big\{ \Big(\frac{1}{2} a f^{1/2}\dot t - \frac{1}{4a} f^{3/2} \dot t^3 \nonumber\\
	& - \frac{5}{4a} f^{-1/2} \dot r^2 \dot t \Big) \frac{1-f}{r} \nonumber\\
	& + \frac{1}{2a} f^{3/2} r \big( \dot t \dot \theta^2 + \sin^2 \theta \, \dot t \dot \phi^2 \big) \nonumber \\
	& + \, \dot r \Big[ \frac{1}{a}\Big( -f^{-3/2} + \frac{1}{2}f^{-1/2} \Big) \dot t \dot r \frac{1-f}{r} \nonumber\\
	& + \, \frac{1}{2a^3} \Big( - \big\{ f^{1/2} \dot t^3 \dot r + f^{-3/2} \dot t \dot r^3 \big\} \frac{1-f}{r} \nonumber\\
	& + 2f^{1/2}r \big\{ \dot t \dot r \dot \theta^2 + \sin^2 \theta \, \dot t \dot r \dot \phi^2 + \frac{r}{2} \sin 2\theta \, \dot t \dot \theta \dot \phi^2 \big\} \Big) \Big]\Big\} \label{path1}
\end{align}
\begin{align}
	&\ddot \theta + \frac{2}{r} \dot \theta \dot r - \frac{1}{2}\sin 2\theta \, \dot \phi^2 \nonumber\\
	& = -\frac{z}{1+z} \tilde A_4 \Big\{ \frac{1}{a} \frac{1}{r} \Big(\frac{1}{4} f^{1/2} \sin 2\theta \, \dot t \dot \phi^2 + 2\dot t \dot r \dot \theta \Big) \nonumber \\
	& + \, \dot \theta \Big[ \frac{1}{a}\Big( -f^{-3/2} + \frac{1}{2}f^{-1/2} \Big) \dot t \dot r \frac{1-f}{r} \nonumber\\
	& + \, \frac{1}{2a^3} \Big( - \big\{ f^{1/2} \dot t^3 \dot r + f^{-3/2} \dot t \dot r^3 \big\} \frac{1-f}{r} \nonumber\\
	& + 2f^{1/2}r \big\{ \dot t \dot r \dot \theta^2 + \sin^2 \theta \, \dot t \dot r \dot \phi^2 + \frac{r}{2} \sin 2\theta \, \dot t \dot \theta \dot \phi^2 \big\} \Big) \Big]\Big\} \label{path2}
\end{align}
\begin{align}
	&\ddot \phi + \frac{2}{r} \dot \phi \dot r + 2\cot \theta \, \dot \theta \dot \phi \nonumber\\
	& = -\frac{z}{1+z} \tilde A_4 \Big\{ \frac{1}{a} \Big( \frac{1}{r} \dot t \dot r \dot \phi + \cot\theta \, \dot t \dot \theta \dot \phi \Big) \nonumber\\
	& + \,\dot \phi \Big[ \frac{1}{a}\Big( -f^{-3/2} + \frac{1}{2}f^{-1/2} \Big) \dot t \dot r \frac{1-f}{r} \nonumber\\
	& + \, \frac{1}{2a^3} \Big( - \big\{ f^{1/2} \dot t^3 \dot r + f^{-3/2} \dot t \dot r^3 \big\} \frac{1-f}{r} \nonumber\\
	& + 2f^{1/2}r \big\{ \dot t \dot r \dot \theta^2 + \sin^2 \theta \, \dot t \dot r \dot \phi^2 + \frac{r}{2} \sin 2\theta \, \dot t \dot \theta \dot \phi^2 \big\} \Big) \Big]\Big\} \label{path3}
\end{align}
with $f = 1-\frac{R_S}{r}$, $R_S$ the Schwarzschild radius.

As an application for our model we present 
a numerical solution (using the differential equation solver of Mathematica) of the timelike path equations (\ref{path0}-\ref{path3}) for an appropriate choice of parameters and initial values. For this application we consider 
that $\theta$=$\frac{\pi}{2}$ while $ t,r $ and $\phi$ are 
the variables. Below we present two figures which 
clarify the difference between 
Schwarzschild-Finsler-Randers (S-F-R) paths (red line) and General 
Relativity (GR) geodesics (blue line). Notice, that the full analysis 
of timelike paths 
as well as the applications to Astrophysics will be studied in a forthcoming 
paper.

\begin{figure}[h]
	\centering
	\includegraphics[scale=.4]{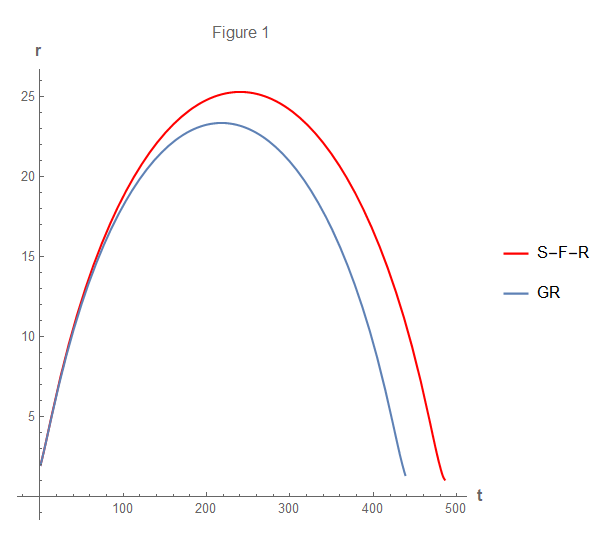}
	\caption{This is an $ r,t $ graph of the timelike paths that we find using our theoretical model Schwarzschild-Finsler-Randers (S-F-R) in comparison to the geodesics of General Relativity (GR).}
\end{figure}

From the figures we observe that in the case of 
S-F-R model the timelike path reaches a higher maximum distance which 
is somewhat larger than the path provided by General Relativity. 
We notice that the time it takes for the S-F-R path to reach 
the Schwarzschild radius is more than in GR.
From these observations we see that in our model the maximum 
radial distance of the orbit is greater than that of GR and 
also the rate at which the particle falls is slower.
As expected, this deviation from the GR geodesic is produced from the right hand side of (\ref{path0}-\ref{path3}),  
where the extra terms in the S-F-R model act as a force that opposes gravity. Since the extra terms of the rhs of (\ref{path0}-\ref{path3}) 
are taken to be small the corresponding 
deviation from the GR geodesic is relatively small.

\begin{figure}[h]
	\centering
	\includegraphics[scale=.4]{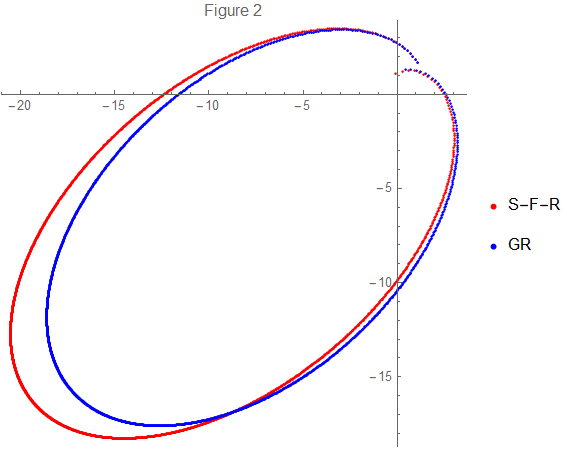}
	\caption{This is an $r,\phi$ polar graph of the timelike paths that we find using our theoretical Schwarzschild-Finsler-Randers (S-F-R; red curve) 
		model in comparison to the geodesics of General Relativity (GR; blue line).}
	
\end{figure}

\subsection{Spacelike paths}
For completeness, we will find the spacelike paths from \eqref{genpath2} following the same procedure as above. From \eqref{spacelike_v} we get
\begin{equation}\label{wnl_space}
	w_{\nu\lambda} = \, g_{\lambda\rho} A_\nu u^\rho + g_{\nu\rho} A_\lambda u^\rho + \, \left( g_{\nu\lambda} - g_{\nu\sigma} g_{\lambda\tau} u^\sigma u^\tau \right) A_\rho u^\rho 
\end{equation}
where, as before, we have lowered down $w_{\alpha\beta}$ to the horizontal space using the generalized Kronecker symbols and we have set $ u^\nu = y^\nu/a $ where $a = \sqrt{g_{\mu\nu}y^\mu y^\nu}$. Following the same steps as for the timelike section of $g_{\mu\nu}$ and taking into account that $g_{\mu\nu}u^\mu u^\nu = 1$, \eqref{genpath2} gives
\begin{align}\label{spacelike_path}
	&\ddot x^\mu + \gamma^\mu_{\kappa\lambda} \dot x^\kappa \dot x^\lambda \nonumber \\
	& = - \frac{z}{1+z} \Big\{ - a g^{\mu\nu} \left(\partial_\nu A_\kappa - \partial_\kappa A_\nu \right) \dot x^\kappa \nonumber\\
	& + \frac{1}{a} \Big[A^\nu \left(\partial_\kappa g_{\nu\lambda} - \frac{1}{2}\partial_\nu g_{\kappa\lambda} \right) + \partial_\kappa A_\lambda \Big] \dot x^\mu \dot x^\kappa \dot x^\lambda \nonumber\\
	& + \, \frac{1}{a} \Big( -\frac{1}{4}g^{\mu\nu} A_\kappa \partial_\nu g_{\sigma\lambda} + g^{\mu\nu} A_\kappa \partial_\lambda g_{\sigma\nu} \Big) \dot x^\sigma \dot x^\kappa \dot x^\lambda \nonumber \\
	& - \, \frac{5}{2a^3} A_\lambda \partial_\kappa g_{\sigma\tau} \dot x^\sigma \dot x^\tau \dot x^\mu \dot x^\kappa \dot x^\lambda \Big\}
\end{align}
Substituting to \eqref{spacelike_path} the solution \eqref{Asolution} we get the explicit form of the spacelike paths components for $r>R_S$:
\begin{align}
	&\ddot t + \frac{1-f}{rf}\dot r \dot t \nonumber\\
	& = -\frac{z}{1+z} \tilde A_4 \Big\{ \Big(-\frac{1}{2}af^{-3/2} \dot r + \frac{1}{a}f^{-1/2} \dot r \dot t^2 \Big)\frac{1-f}{r} \nonumber\\
	& +  \dot t \Big[ \frac{1}{a}\Big( -f^{-3/2} + \frac{1}{2}f^{-1/2} \Big) \dot t \dot r \frac{1-f}{r} \nonumber\\
	& -  \frac{5}{2a^3} \Big( - \big\{ f^{1/2} \dot t^3 \dot r + f^{-3/2} \dot t \dot r^3 \big\} \frac{1-f}{r} \nonumber\\
	& + 2f^{1/2}r \big\{ \dot t \dot r \dot \theta^2 + \sin^2 \theta \, \dot t \dot r \dot \phi^2 + \frac{r}{2} \sin 2\theta \, \dot t \dot \theta \dot \phi^2 \big\} \Big) \Big]\Big\}
\end{align}
\begin{align}
	&\ddot r + \frac{f(1-f)}{2r} \dot t^2 - \frac{1-f}{2rf} \dot r^2 - rf \big( \dot \theta^2 + \sin^2 \theta \dot  \phi^2 \big) \nonumber\\
	& = -\frac{z}{1+z} \tilde A_4 \Big\{ \Big(-\frac{1}{2} a f^{1/2}\dot t + \frac{1}{4a} f^{3/2} \dot t^3 \nonumber\\
	& - \frac{3}{4a} f^{-1/2} \dot r^2 \dot t \Big) \frac{1-f}{r} \nonumber\\
	& - \frac{1}{2a} f^{3/2} r \big( \dot t \dot \theta^2 + \sin^2 \theta \, \dot t \dot \phi^2 \big) \nonumber \\
	& + \dot r \Big[ \frac{1}{a}\Big( -f^{-3/2} + \frac{1}{2}f^{-1/2} \Big) \dot t \dot r \frac{1-f}{r} \nonumber\\
	& - \frac{5}{2a^3} \Big( - \big\{ f^{1/2} \dot t^3 \dot r + f^{-3/2} \dot t \dot r^3 \big\} \frac{1-f}{r} \nonumber\\
	& + 2f^{1/2}r \big\{ \dot t \dot r \dot \theta^2 + \sin^2 \theta \, \dot t \dot r \dot \phi^2 + \frac{r}{2} \sin 2\theta \, \dot t \dot \theta \dot \phi^2 \big\} \Big) \Big]\Big\}
\end{align}
\begin{align}
	&\ddot \theta + \frac{2}{r} \dot \theta \dot r - \frac{1}{2}\sin 2\theta \, \dot \phi^2 \nonumber\\
	& = -\frac{z}{1+z} \tilde A_4 \Big\{ \frac{1}{a} \frac{1}{r} \Big(-\frac{1}{4} f^{1/2} \sin 2\theta \, \dot t \dot \phi^2 + 2\dot t \dot r \dot \theta \Big) \nonumber \\
	& + \dot \theta \Big[ \frac{1}{a}\Big( -f^{-3/2} + \frac{1}{2}f^{-1/2} \Big) \dot t \dot r \frac{1-f}{r} \nonumber\\
	& - \frac{5}{2a^3} \Big( - \big\{ f^{1/2} \dot t^3 \dot r + f^{-3/2} \dot t \dot r^3 \big\} \frac{1-f}{r} \nonumber\\
	& + 2f^{1/2}r \big\{ \dot t \dot r \dot \theta^2 + \sin^2 \theta \, \dot t \dot r \dot \phi^2 + \frac{r}{2} \sin 2\theta \, \dot t \dot \theta \dot \phi^2 \big\} \Big) \Big]\Big\}
\end{align}
\begin{align}
	& \ddot \phi + \frac{2}{r} \dot \phi \dot r + 2\cot \theta \, \dot \theta \dot \phi \nonumber\\
	& = -\frac{z}{1+z} \tilde A_4 \Big\{ \frac{1}{a} \Big( \frac{1}{r} \dot t \dot r \dot \phi + \cot\theta \, \dot t \dot \theta \dot \phi \Big) \nonumber\\
	& + \dot \phi \Big[ \frac{1}{a}\Big( -f^{-3/2} + \frac{1}{2}f^{-1/2} \Big) \dot t \dot r \frac{1-f}{r} \nonumber\\
	& - \, \frac{5}{2a^3} \Big( - \big\{ f^{1/2} \dot t^3 \dot r + f^{-3/2} \dot t \dot r^3 \big\} \frac{1-f}{r} \nonumber\\
	& + 2f^{1/2}r \big\{ \dot t \dot r \dot \theta^2 + \sin^2 \theta \, \dot t \dot r \dot \phi^2 + \frac{r}{2} \sin 2\theta \, \dot t \dot \theta \dot \phi^2 \big\} \Big) \Big]\Big\}
\end{align}

\section{Conclusion}

In this paper we derived for the first time 
the gravitational field as a solution of the spherically 
symmetric Schwarzschild-Randers and Schwarzschild-De Sitter-Randers metric.
In this framework we used generalized Einstein field equations on the tangent bundle of a spacetime with zero horizontal energy-momentum tensor in which we get more degrees of freedom.
In addition, we specified an appropriate timelike covector  which plays a significant role in this theory differentiating our model from the traditional Schwarzschild one, giving an intrinsic anisotropic character to the ordinary Schwarzschild metric as well as for the particle paths. Moreover, we studied the correlation of the Schwarzschild-De Sitter model with the Randers one as becomes apparent from \eqref{Asolution2}. We also studied the forms of paths in our spacetime and we obtained more generalized forms than the ordinary geodesic paths of the classical Schwarzschild spacetime. In this context, 
we provided some numerical solutions of the timelike 
paths of our model and we found small but not negligible deviations from 
General Relativity. This difference can been seen 
as a result of the local 
anisotropy which creates an effective force that affects the 
corresponding geodesics.

It is obvious that when the covector $A_\gamma$ of our theory vanishes then we recover the ordinary form of a Schwarzschild metric and Schwarzschild-De Sitter metric respectively and their derived geodesics.

Such an approximation can be considered compatible with some current observational data and parameters with anisotropic character in cosmological models of Schwarzschild-Randers and Schwarzschild-De Sitter-Randers spacetimes. 
These features mean that Finsler-Randers gravity
can be interesting at the astrophysical level.
This study will be the goal of our next work.

\appendix

\section{Basic structures on the Lorentz tangent bundle}\label{sec:preliminaries}

The 8-dimensional Lorentz tangent bundle $TM$ is equipped with local coordinates $\{\mathcal{U}^A \} = \{x^\mu,y^\alpha\}$ where $x^\mu$ are the local coordinates on the base manifold $M$ around $\pi(\sigma)$, $\sigma \in TM$, and $y^\alpha$ are the coordinates on the fiber. The range of values for the indices is $\kappa,\lambda,\mu,\nu,\ldots = 0,\ldots,3$ and $\alpha,\beta,\ldots,\theta = 4,\ldots,7$.

The adapted basis on the total space $TTM$ is defined as $\{E_A\} = \,\{\delta_\mu,\dot\partial_\alpha\} $ where
\begin{equation}
	\delta_\mu = \dfrac{\delta}{\delta x^\mu}= \pder{}{x^\mu} - N^\alpha_\mu(x,y)\pder{}{y^\alpha} \label{delta x}
\end{equation}
and
\begin{equation}
	\dot \partial_\alpha = \pder{}{y^\alpha}
\end{equation}
where $N^\alpha_\mu$ are the components of a nonlinear connection. The curvature of the nonlinear connection is defined as
\begin{equation}\label{Omega}
	\Omega^\alpha_{\nu\kappa} = \dder{N^\alpha_\nu}{x^\kappa} - \dder{N^\alpha_\kappa}{x^\nu}
\end{equation}
The nonlinear connection induces a split of the total space $TTM$ into a horizontal distribution $T_HTM$ and a vertical distribution $T_VTM$. The above-mentioned split is expressed with the Whitney sum:
\begin{equation}
	TTM = T_HTM \oplus T_VTM.
\end{equation}
The horizontal distribution or h-space is spanned by $\delta_\mu$, while the vertical distribution or v-space is spanned by $\pdot \alpha$. Under a local coordinate transformation on the base manifold, the adapted basis vectors transform as:
\begin{equation}\label{h basis transformation}
	\delta_{\mu'} = \pder{x^\mu}{x^{\mu'}}\delta_\mu \quad,\quad \pdot{\alpha'} = \pder{x^\alpha}{x^{\alpha'}}\pdot{\alpha}.
\end{equation}
with $x^\alpha = \tilde \delta^\alpha_\mu x^\mu $ \footnote{The generalized Kronecker symbols are defined as: $\tilde \delta_\alpha^\mu = \tilde\delta^\alpha_\mu = 1$ for $a=\mu+4$ and equal to zero otherwise.}. The adapted dual basis of the adjoint total space $T^*TM$ is $ \{E^A \}  = \{\de x^\mu, \delta y^\alpha\}$ with the definition
\begin{equation}
	\delta y^\alpha = \mathrm{d}y^\alpha + N^\alpha_\nu\mathrm{d}x^\nu .\label{delta y}
\end{equation}
The transformation rule for $ \{\de x^\mu, \delta y^\alpha\} $ is:
\begin{equation}
	\de x^{\mu'} = \pder{x^{\mu'}}{x^\mu}\de x^\mu \quad,\quad \delta y^{\alpha'} = \pder{x^{\alpha'}}{x^\alpha}\delta y^\alpha
\end{equation}
The bundle $TM$ 

In this work, we consider a distinguished connection ($d-$connection) $ {D} $ on $TM$. This is a linear connection with coefficients $\{\Gamma^A_{BC}\} = \{L^\mu_{\nu\kappa}, L^\alpha_{\beta\kappa}, C^\mu_{\nu\gamma}, C^\alpha_{\beta\gamma} \} $ which preserves by parallelism the horizontal and vertical distributions:
\begin{align}
	{D_{\delta_\kappa}\delta_\nu = L^\mu_{\nu\kappa}(x,y)\delta_\mu}\, \quad &,\quad D_{\pdot{\gamma}}\delta_\nu = C^\mu_{\nu\gamma}(x,y)\delta_\mu \label{D delta nu} \lin
	{D_{\delta_\kappa}\pdot{\beta} = L^\alpha_{\beta\kappa}(x,y)\pdot{\alpha}} \quad &, \quad D_{\pdot{\gamma}}\pdot{\beta} = C^\alpha_{\beta\gamma}(x,y)\pdot{\alpha} \label{D partial b}
\end{align}
From these, the definitions for partial covariant differentiation follow as usual, e.g. for $X \in TTM$ we have the definitions for covariant h-derivative
\begin{equation}
	X^A_{|\nu} \equiv D_\nu\,X^A \equiv \delta_\nu X^A + L^A_{B\nu}X^B \label{vector h-covariant}
\end{equation}
and covariant v-derivative
\begin{equation}
	X^A|_\beta \equiv D_\beta\,X^A \equiv \dot{\partial}_\beta X^A + C^A_{B\beta}X^B \label{vector v-covariant}
\end{equation}

A $d-$connection can be uniquely defined given that the following conditions are satisfied:
\begin{itemize}
	\item The $d-$connection is metric compatible
	\item Coefficients $L^\mu_{\nu\kappa}, L^\alpha_{\beta\kappa}, C^\mu_{\nu\gamma}, C^\alpha_{\beta\gamma} $ depend solely on the quantities $g_{\mu\nu}$, $v_{\alpha\beta}$ and $N^\alpha_\mu$
	\item Coefficients $L^\mu_{\kappa\nu}$ and $ C^\alpha_{\beta\gamma} $ are symmetric on the lower indices, i.e.  $L^\mu_{[\kappa\nu]} = C^\alpha_{[\beta\gamma]} = 0$
\end{itemize}
We use the symbol $\mathcal D$ instead of $D$ for a connection satisfying the above conditions, and call it a canonical and distinguished $d-$connection. Metric compatibility translates into the conditions:
\begin{equation}
	\mathcal D_\kappa\, g_{\mu\nu} = 0, \quad \mathcal D_\kappa\, v_{\alpha\beta} = 0, \quad\mathcal D_\gamma\, g_{\mu\nu} = 0, 
	\quad\mathcal D_\gamma\, v_{\alpha\beta} = 0.
\end{equation}
The coefficients of canonical and distinguished $d-$ connection are
\begin{align}
	L^\mu_{\nu\kappa} & = \frac{1}{2}g^{\mu\rho}\left(\delta_kg_{\rho\nu} + \delta_\nu g_{\rho\kappa} - \delta_\rho g_{\nu\kappa}\right) \label{metric d-connection 1}  \\
	L^\alpha_{\beta\kappa} & = \dot{\partial}_\beta N^\alpha_\kappa + \frac{1}{2}v^{\alpha\gamma}\left(\delta_\kappa v_{\beta\gamma} - v_{\delta\gamma}\,\dot{\partial}_\beta N^\delta_\kappa - v_{\beta\delta}\,\dot{\partial}_\gamma N^\delta_\kappa\right) \label{metric d-connection 2}  \\
	C^\mu_{\nu\gamma} & = \frac{1}{2}g^{\mu\rho}\dot{\partial}_\gamma g_{\rho\nu} \label{metric d-connection 3} \\
	C^\alpha_{\beta\gamma} & = \frac{1}{2}v^{\alpha\delta}\left(\dot{\partial}_\gamma v_{\delta\beta} + \dot{\partial}_\beta v_{\delta\gamma} - \dot{\partial}_\delta v_{\beta\gamma}\right). \label{metric d-connection 4}
\end{align}

Curvature and torsion in $TM$ can be defined as multi-linear maps:
\begin{equation}
	\mathcal{R}(X,Y)Z = [\mathcal{D}_X,\mathcal{D}_Y]Z - \mathcal{D}_{[X,Y]}Z \label{Riemann tensor TM}
\end{equation}
and
\begin{equation}
	\mathcal{T}(X,Y) = \mathcal{D}_XY - \mathcal{D}_YX - [X,Y], \label{torsion TM}
\end{equation}
where $X,Y,Z \in TTM$.
We use the definitions
\begin{align}
	\mathcal{R}(\delta_\lambda,\delta_\kappa)\delta_\nu = R^\mu_{\nu\kappa\lambda}\delta_\mu \label{R curvature components} \lin
	\mathcal{R}(\pdot{\delta},\pdot{\gamma})\pdot{\beta} = S^\alpha_{\beta\gamma\delta}\dot{\partial}_\alpha \label{S curvature components}
\end{align}

\begin{align}
	\mathcal{T}(\delta_\kappa,\delta_\nu) = & \mathcal{T}^\mu_{\nu\kappa}\delta_\mu + \mathcal{T}^\alpha_{\nu\kappa}\pdot{\alpha} \label{torsion components 1} \lin
	\mathcal{T}(\pdot{\gamma},\pdot{\beta}) = & \mathcal{T}^\mu_{\beta\gamma}\delta_\mu + \mathcal{T}^\alpha_{\beta\gamma}\pdot{\alpha} \label{torsion components 3}
\end{align}
The h-curvature tensor of the $d-$connection in the adapted basis and the corresponding h-Ricci tensor have, respectively, the components
\begin{align}
	R^\mu_{\nu\kappa\lambda} =  \delta_\lambda L^\mu_{\nu\kappa} - \delta_\kappa L^\mu_{\nu\lambda} + L^\rho_{\nu\kappa}L^\mu_{\rho\lambda} - L^\rho_{\nu\lambda}L^\mu_{\rho\kappa} + C^\mu_{\nu\alpha}\Omega^\alpha_{\kappa\lambda} \label{R coefficients 1}
\end{align}
\begin{align}
	R_{\mu\nu} = R^\kappa_{\mu\nu\kappa} = & \, \delta_\kappa L^\kappa_{\mu\nu} - \delta_\nu L^\kappa_{\mu\kappa} + L^\rho_{\mu\nu}L^\kappa_{\rho\kappa} \nonumber
	- L^\rho_{\mu\kappa}L^\kappa_{\rho\nu} \nonumber\\
	& + C^\kappa_{\mu\alpha}\Omega^\alpha_{\nu\kappa} \label{d-ricci 1}
\end{align}
The v-curvature tensor of the $d-$connection in the adapted basis and the corresponding v-Ricci tensor have, respectively, the components
\begin{align}
	S^\alpha_{\beta\gamma\delta} & = \pdot{\delta} C^\alpha_{\beta\gamma} - \pdot{\gamma}C^\alpha_{\beta\delta} + C^\epsilon_{\beta\gamma}C^\alpha_{\epsilon\delta} - C^\epsilon_{\beta\delta}C^\alpha_{\epsilon\gamma} \label{S coefficients 2} \lin
	S_{\alpha\beta} & = S^\gamma_{\alpha\beta\gamma} = \pdot{\gamma}C^\gamma_{\alpha\beta} - \pdot{\beta}C^\gamma_{\alpha\gamma} + C^\epsilon_{\alpha\beta}C^\gamma_{\epsilon\gamma} - C^\epsilon_{\alpha\gamma}C^\gamma_{\epsilon\beta}. \label{d-ricci 4}
\end{align}
The generalized Ricci scalar curvature in the adapted basis is defined as
\begin{equation}
	\R = g^{\mu\nu}R_{\mu\nu} + v^{\alpha\beta}S_{\alpha\beta} = R+S. \label{bundle ricci curvature}
\end{equation}
where
\begin{align}
	R=g^{\mu\nu}R_{\mu\nu} \quad,\quad
	S=v^{\alpha\beta}S_{\alpha\beta} \label{hv ricci scalar}
\end{align}
In the main text, we use the more convenient definitions
\begin{align}\label{Rmnline}
	\overline R_{\mu\nu} = & \, R_{\mu\nu} - C^\kappa_{\mu\alpha}\Omega^\alpha_{\nu\kappa} = \delta_\kappa L^\kappa_{\mu\nu} - \delta_\nu L^\kappa_{\mu\kappa} + L^\rho_{\mu\nu}L^\kappa_{\rho\kappa} \nonumber\\
	& - L^\rho_{\mu\kappa}L^\kappa_{\rho\nu}
\end{align}
\begin{equation}\label{Rline}
	\overline R = g^{\mu\nu}\overline R_{\mu\nu}
\end{equation}

\section{Calculation of $A_\gamma$}\label{Acalculation}

\subsection{Solution for the Schwarzschild-Randers spacetime}
In order to calculate $A_{\gamma}$ we will give values to $\mu,\nu,\gamma$ of \eqref{fullexp} and solve the resulting equations.
For $\mu=0, \nu=0$ and $\gamma=4$ we get :
\begin{equation}
	\partial^{2}_{0} A_{4} + \frac{1}{2}f\partial_{1}(-f)\left(\partial_{1}A_{4}+\frac{1}{2f}A_{4}\partial_{1}(-f)\right)=0,
\end{equation}
where we have set $f=1-\frac{R_s}{r}$.
After some calculations and by separation of variables $A_{4}=R_4(r)T_4(t)$ we get two equations:
\begin{equation}\label{teq0}
	\partial_{0}T_4(t)=-c^{2}_{(4)}T_4(t)
\end{equation}
\begin{equation}\label{req0}
	\partial_{1}R_4(r)=\frac{1-f}{2rf}R_4(r)-c^{2}_{(4)}\frac{2r}{f(1-f)}
\end{equation}
where $c^{2}_{(4)}$ is the separation constant.
For $\mu=0, \nu=1$ and $\gamma=4$ we get
\begin{equation}
	\partial_{0}\partial_{1}A_{4} +\frac{1}{f}\partial_{1}(-f)\partial_{0}A_{4} = 0.
\end{equation}
After rearranging the terms and again separating variables, for $\partial_0T_4 \neq0$ we have 
\begin{equation}
	\partial_{1}R_4(r) = \frac{\partial_{1}f}{f}R_4(r)
\end{equation}
So we get $R_4(r)=k_4f(r)$ with $k_4$ being a constant resulting from the integration. By substituting this to \eqref{req0} we find that the separation constant $c^{2}_{(4)}$ must be zero. That means that in order to satisfy \eqref{teq0} and \eqref{req0}, $T_4(t)$ must be constant and $R_4(r)=\tilde{R}_4f^{1/2}(r)$ with $\tilde{R}_4$ a constant.
By calculating the remaining equations for $\mu=2, \nu=2, \gamma=4$ and $\mu=3,\nu=3,\gamma=4$ we find that $A_{4}$ has no dependence on $\theta$ or $\phi$. Therefore, we end up with 
\begin{equation}
	A_{4}=\tilde A_{4} |f|^{1/2}(r)
\end{equation}
with $\tilde A_{4} $ being a constant.
For $\mu=0, \nu=0, \gamma=5$ we get 
\begin{equation}
	\partial^{2}_{0}A_{5} - \frac{f(1-f)}{2r}\left(\partial_{1}A_{5} - \frac{1}{2}A_{5}f\partial_{1}(f^{-1})\right)=0.
\end{equation}
After calculations and by separating variables like before we end up with equations
\begin{equation}\label{teq1}
	\partial^{2}_{0}T_5(t)=-c^{2}_{(5)}T_5(t)
\end{equation}
\begin{equation}\label{req1}
	\partial_{1}R_5(r)=-\frac{1-f}{2rf}R_5(r)-c^{2}_{(5)}\frac{2r}{f(1-f)}
\end{equation}
with $c^{2}_{(5)}$ being the separation constant.
For $\mu=0, \nu=1, \gamma=5$ we get
\begin{equation}
	\partial_{0}\partial_{1} A_{5}-\frac{1}{2}f\partial(f^{-1})\partial_{0}A_{5}-\frac{1}{2}(-f^{-1})\partial_{1}(-f)\partial_{0}A_{5}=0
\end{equation}
After calculations we end up with $c_{(5)}=0$ and by substitution to \eqref{teq1} and \eqref{req1} we find 
\begin{equation}
	R_5(r)=k_5|f|^{-1/2}
\end{equation}
with $k_5$ being a constant.
Also, like before, if we calculate the $\mu=2,\nu=2,\gamma=5$ and $\mu=3,\nu=3,\gamma=5$ we get no dependence on $\theta$ and $\phi$.
Therefore, we find
\begin{equation}
	A_{5}=\tilde A_{5} |f|^{-1/2}(r)
\end{equation}
with $\tilde A_{5} $ a constant of integration. If we put this solution in the $\mu=1, \nu=1$ equation, we get $\tilde A_5 = 0$.
For $\mu=0,\nu=0,\gamma=6$ we separate variables like before $A_{6}=R_6(r)T_6(t)$ and we get two equations:
\begin{equation}\label{req2}
	\partial_{1}R_6(r)=\frac{R_6(r)}{r}-c^{2}_{(6)}\frac{2r}{f(1-f)}
\end{equation}
\begin{equation}\label{teq2}
	\partial^{2}_{0}T_{6}(t)=-c^{2}_{(6)}T_{6}(t)
\end{equation}
with $c^{2}_{(6)}$ the separation constant.
For $\mu=0,\nu=1,\gamma=6$ like before we find that $c_{(6)}=0$ and by substitution to \eqref{req2} we find that $R_6(r)=\tilde R_{6} r$ with $\tilde R_{6} $ a constant of integration. We set $\tilde R_{6} $ to zero to keep our solution finite at infinity, so we end up with $A_{6}=0$.
For $A_{7}$ we set $\mu=0,\nu=0,\gamma=7$ and we find the same equations as for $A_{6}$. That leads to $A_{7}=0$ as well.

To sum up, we have found the following solution for $A_\gamma$ from equation \eqref{fullexp}:
\begin{equation}
	A_\gamma(x) = \left[ \tilde A_4 \left|1-\frac{R_S}{r} \right|^{1/2}, 0, 0, 0 \right]
\end{equation}
with $\tilde A_4 $ a constant. 

\subsection{Solution for the Schwarzschild-De Sitter-Randers spacetime}

We will modify the solution \eqref{Asolution} and see if it satisfies \eqref{fullexp} for the metric \eqref{Schwarzchild-desitter}. An obvious ansatz is to replace the term $ 1-\frac{R_S}{r} $ in \eqref{Asolution} with $ 1-\frac{R_S}{r} - \frac{\Lambda}{3}r^2 $. Doing this we get
\begin{equation}
	A_\gamma(x) = \left[ \tilde A_4 \left|1-\frac{R_S}{r} - \frac{\Lambda}{3}r^2 \right|^{1/2}, 0, 0, 0 \right]
\end{equation}
which is verified to be a solution of \eqref{fullexp} for the metric \eqref{Schwarzchild-desitter}.

\section*{Acknowledgments}
We would like to thank the unknown referees for their valuable comments and suggestions which have helped improve our work. This research is co-financed by Greece and the European Union (European Social Fund-ESF) through the Operational Programme ``Human Resources Development, Education and Lifelong Learning'' in the context of the project ``Strengthening Human Resources Research Potential via Doctorate Research'' (MIS-5000432), implemented by the State Scholarships Foundation (IKY).

\end{document}